\documentclass[aps,epsf,preprint,prl,showpacs]{revtex4-1}
\usepackage{graphicx}
\begin{document}
\bibliographystyle{apsrev}

\title{Electrically generated surface plasmons by electroluminescence of individual carbon nanotube field effect transistor}

\author{P. Rai$^\dagger$, N. Hartmann$^\ddagger$,
J. Berthelot$^\dagger$, J. Arocas$^\dagger$ G. Colas des Francs$^\dagger$, A. Hartschuh$^\ddagger$ and A. Bouhelier$^{\dagger*}$}
\affiliation{$^\dagger$Laboratoire Interdisciplinaire Carnot de
Bourgogne, CNRS-UMR 6303, Universit\'{e} de Bourgogne, 21078 Dijon,
France\\$^\ddagger$Department Chemie and CeNS, Ludwig-Maximilians-Universit\"at, M\"unchen 81377, Germany}

\begin{abstract}
We demonstrate the realization of an electrically-driven integrated source of surface plasmon polaritons. Light-emitting individual single-walled carbon nanotube field effect transistors were fabricated in a plasmonic-ready platform. The devices were operated at ambient condition to act as an electroluminescence source localized near the contacting gold electrodes. We show that photon emission from the semiconducting channel can couple to propagating surface plasmons developing in the electrical terminals. Momentum-space spectroscopy suggests that excited plasmon modes are bound to the metal-glass interface. Our results underline the high degree of compatibility between state-of-the art nano-optoelectronic devices and plasmonic architectures.
\end{abstract}


 \maketitle

Plasmonics is emerging as an alternative technology to satisfy constrains of miniaturization of optical devices down to sub-wavelength sizes~\cite{barnes03,brongersma06,ozbay06}. This on-chip technology utilizes the unique properties of surface plasmon polariton (SPP) to transmit optical and electrical signals through the same metal-based circuitry. A complete plasmonic platform requires the development of elementary building blocks with specific functionalities including routing and modulating SPP, but also on-chip excitation and detection~\cite{brongersma06,ebbesen08,DragomanPQE08}. The vast majority of fabricated plasmonic prototypes rely on an optical excitation of SPPs, typically involving laser sources and macroscopic coupling elements~\cite{bouhelier05OL,Hohenau05}. Despite optimized SPP coupling efficiencies, these external sources can hardly be integrated on the nanoscale and novel routes for in/out connections are needed. An electrical excitation and detection of SPPs offers dramatic advantages over standard coupling schemes. Electrical terminals can be readily integrated offering thus a hybrid technology to interface miniature electronic devices with a plasmonic architecture~\cite{polman10,vandorpe10,brongersma12NL,Bharadwaj11,krenn08}. 

In this context, the unique opto-electronic properties of carbon nanotubes (CNTs)~\cite{avouris08,avouris05science,avouris04} could provide a novel approach towards scaling SPP nano-sources in complex interconnect systems. Electroluminescence (EL) from single-walled nanotubes (SWNTs) can be achieved through different mechanisms, such as impact excitation~\cite{avouris05science}, recombination of carriers during ambipolar transport~\cite{avouris04}, phonon-assisted process~\cite{essig10}, and recombination from thermally populated electronic states~\cite{freitag04}. SWNTs feature well-defined electronic resonances in the visible to near-infrared spectral range and sustain extremely high current densities~\cite{avouris05science,Rao97}. Bright, stable and localized infrared emission was observed from individual SWNT field effect transistors (SWNTFETs) demonstrating their usability as nano-scale light sources~\cite{avouris05science,avouris04}. SWNTs could thus offer a new integration route enabling a local plasmon excitation in a circuitry combining both electrical wiring and radiation source in a single unit superseding complex fabrication procedures.  

In this article we demonstrate coupling of the radiation emitted by an individual electrically-driven SWNTFET to SPPs developing in waveguide-like electrodes. EL emission from the device is obtained by an impact excitation process under unipolar transport of carriers in the semiconducting channel. Theoretical calculation of SPP propagation length and experimental observation of momentum-space radiation patterns confirm the bound nature of the excited SPP mode.

\begin{figure}[t]
\includegraphics[width=8.5cm]{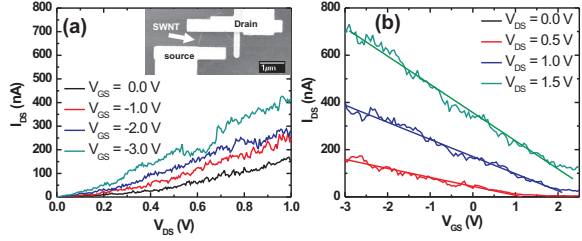}
\caption{\label{figure1} (Color online). (a) The output and (b) transfer characteristics of a SWNTFET. A straight line fitting follows the transport equation of the device. The inset in (a) shows a SEM image of the SWNTFET fabricated on a glass cover slip.}
\end{figure}

SWNTFETs were fabricated by standard electron beam lithography and lift off processes on a glass cover slip. Electrical gating was insured by evaporating a 30~nm thick conductive layer of transparent indium tin oxide (ITO). A 250~nm thick layer of SiO$_{2}$ was then deposited on the ITO to act as a dielectric spacer. A solution of high pressure carbon monoxide synthesized SWNTs~\cite{hartmann12,Rai:12} were randomly deposited on the prepared ITO-gated substrate. Nanotubes were located by scanning electron microscopy (SEM) before lithography of the electrical source and drain contacts. The contacts were formed by depositing 5~nm of Pd followed by 45~nm of Au. Pd was employed to reduce the Schottky barrier for majority carriers in the nanotube~\cite{avouris08,avouris05science} while Au was used for its compatibility to simultaneously transport electrons and SPPs. The inset of Fig.~\ref{figure1}(a) shows a SEM image of a $\sim$0.8~$\mu$m long SWNTFET. The contrast was saturated to distinguish the nanotube from the electrodes. Drain was maintained at ground potential. The diameter of the nanotube is $\sim$0.75~nm as measured with atomic force microscopy. The output characteristics ($I_{DS}$-$V_{DS}$) of the device for several gate bias ($V_{GS}$) is presented in Fig.~\ref{figure1}(a). The drain current ($I_{DS}$) is increasing with negative gate bias. Figure~\ref{figure1}(b) displays the transfer characteristics ($I_{DS}$-$V_{GS}$) for different $V_{DS}$ showing a $p$-channel metal-oxide-semiconductor field-effect transistor behavior (FET). The ON/OFF ratio of the transistor is $\sim$50. All the nanotube devices fabricated on glass cover slips (around 50) were behaving like FETs with majority carriers dominated by holes. Several mechanisms were proposed to explain $p$-type conduction in SWNTFETs, including adsorption of oxygen~\cite{Sumanasekera00}, doping introduced during processing of the nanotubes~\cite{martel98}, or Schottky barrier formation at the nanotube-metal contacts~\cite{avouris05science,avouris02}. The latter is playing a central role in switching off SWNTFETs by modulating the contact resistance rather than the channel conductance~\cite{avouris02}. The origin of $p$-type conduction in our device is mainly due Pd/Au contacts and operation at ambient condition~\cite{avouris08,avouris05science}.

\begin{figure}[t]
\includegraphics[width=8.5cm]{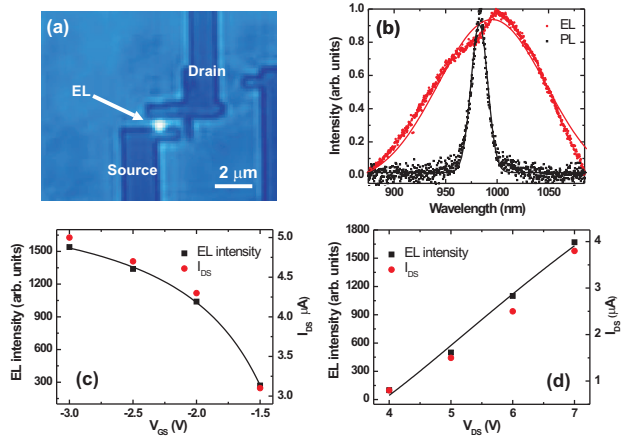}
\caption{\label{figure2} (Color online). (a) EL activity of the SWNTFET at $V_{GS}$=-3~V, $V_{DS}$=6~V and for an integration time 60~s. (b) Normalized PL and EL spectrum. (c) Plot of EL intensity (black squares) and drain current (red circles) versus gate bias at constant $V_{DS}$=6~V. (d) Plot of EL intensity and drain current versus drain bias at constant $V_{GS}$=-2~V.The solid black lines in (c) and (d) are fits of Eq.\ref{eq1} to the experimental data.}
\end{figure}

We recorded both spatial and spectral distributions of the EL with an inverted optical microscope equipped with 100$\times$, 1.49 numerical aperture (N.A.) oil immersion objective and an imaging spectrometer. Figure~\ref{figure2}(a) shows an overlaid image of the EL emission and the contacting electrode geometry obtained under a weak optical illumination. The EL image was recorded at $V_{DS}$=6~V and $V_{GS}$=-3~V for 60~s. Within the resolution of the microscope ($\sim$500~nm, two pixels of the camera), EL emission is observed over the entire length of the channel. To identify the optical transition responsible for EL emission, we compared it to the photoluminescence (PL) of individual nanotubes from the same colloidal solution placed on the same substrate. This procedure was necessary because the nanotubes imaged by SEM during the device fabrication did not show PL emission~\cite{Rai:12}. The EL (red square) and PL (black circle) spectra are shown together in Fig.~\ref{figure2}(b), and are centered at 995~nm and 982~nm, respectively. The emission correlates with the $E_{11}$ excitonic transition of (6,5) nanotubes. The EL spectra is broadened compared to PL emission probably due to the involvement of various vibrational states excited during electronic transport~\cite{freitag04}. A small red shift in the EL may be caused by drain-induced doping~\cite{freitag09}.

We hypothesize that an impact excitation process is responsible for EL emission because the emission intensity depends on the application of a bias voltage. In short channel SWNTFETs, the combined electric fields applied between source and drain and between source and gate produce a sufficiently high electric field at the metal-nanotube interface to accelerate the carriers ($\sim$MV/cm estimated from the bias voltage applied to the nanotube channel). Thus, holes injected into valence band are accelerated towards the drain by the large electrical potential difference. The accelerated holes may produce singlet excitons by impact excitation, which can decay radiatively~\cite{avouris05science}. Impact excitation requires conserving both energy and momentum. Momentum conservation increases the threshold energy by a factor of 1.5 with respect to the minimum energy necessary for exciton formation~\cite{okuto72}. However, this requirement could be relaxed in SWNTs due to interaction with the substrate and exciton band mixing effects reducing the threshold energy close to the exciton energy~\cite{perebeinos04}.

The threshold electric field ($F_{th}$) for impact excitation is given by $\sim 1.5 E_g/e\lambda_{ph}$, where $E_g$ and $\lambda_{ph}$ are the optical band gap and optical phonon scattering length of the SWNT, respectively. $E_g$ deduced from the EL/PL emission is about 1.25~eV. $\lambda_{ph}$ is approximately 15-20~nm for SWNTs~\cite{javey04}. $F_{th}$ is thus estimated to be 0.9-1.2~MV/cm. The EL emission intensity is depending on the impact excitation rate with an exponent equal to $[-F_{th}/F_{app}]$, where $F_{app}$ is the effective electric field applied to the metal-nanotube interface~\cite{okuto72}.  This electric field can be expressed as $ F_{app}$ = $[\gamma V_{DS}+(V_{GS}-V_T)]/\lambda_{scr}$, where $\gamma$ is the fraction of $V_{DS}$ contributing to the impact excitation rate, $\lambda_{scr}$ is the screening length that represents the length over which the potential drops within the metal-nanotube junction~\cite{franklin09} and $V_T$ is the turn-on voltage for the device. Thus, EL emission intensity ($S_{el}$) is given by the expression:

\begin{equation}
S_{el} \sim \exp[-F_{th} \lambda_{scr}/(\gamma V_{DS}+(V_{GS}-V_T))].
\label{eq1}
\end{equation}

The screening length for impact excitation in SWNTFET devices can be expressed as~\cite{franklin09}  $\lambda_{scr}= \sqrt{ \epsilon_{cnt} d_{cnt} t_{ox}/\epsilon_{ox}}$, where $\epsilon_{cnt}$ and $d_{cnt}$ are the dielectric constant and diameter of the CNT, respectively and $\epsilon_{ox}$ and $t_{ox}$ are the dielectric constant and thickness of the oxide layer, respectively. The screening length is estimated to be $\sim$30~nm by using of $\epsilon_{cnt}$=18~\cite{fagan07}, $d_{cnt}$=0.75~nm, $\epsilon_{ox}$=4 and $t_{ox}$=250~nm. The transport properties of SWNTFET is governed by the equation, $I_{DS}=\mu C/L^2 V_{DS} (V_{GS}-V_T)$, where $\mu$ is the carrier mobility, $C$ is the geometrical capacitance and $L$ is the channel length~\cite{martel98}. The geometrical capacitance of a nanotube is~\cite{Anantram06} $C = 2\pi \epsilon_{ox}/\ln (2 t_{ox}/d_{cnt})$=0.04~aF/nm. $V_T$ $\sim$2.2~V and mobility $\mu$ $\sim$25~cm$^2$V/s for holes are deduced by a straight line fitting of the transfer characteristics with transport equation (Fig.~\ref{figure1}(b))Ò. 

To confirm the mechanism of impact excitation responsible for light emission in our devices, we recorded the emission intensity by varying the bias voltages. Figure~\ref{figure2}(c) shows the plot of $I_{DS}$ and EL emission intensity versus gate bias, keeping $V_{DS}$ constant at 6~V. The EL intensity was obtained by integrating the number of photon counts received by the camera in an area corresponding to the length of the channel. Both EL intensity (black squares) and $I_{DS}$ (red circles) are showing the same dependence on gate bias demonstrating the relationship between the number of carriers in the channel and the EL activity.  The evolution with $V_{GS}$ is well reproduced by Eq.~\ref{eq1} as shown by the solid black line in Fig.~\ref{figure2}(c) with $\gamma$=0.33. The remaining $V_{DS}$ may be attributed to carrier transport and thermal heating in the nanotube. Figure~\ref{figure2}(d) depicts the EL response with drain bias for $V_{GS}$=-2~V (black squares). The current injected in the channel is shown with red circles. Here too, there is an obvious correspondence between the EL activity and the drain-controlled current running through the transistor and that is well described by Eq.~\ref{eq1}  (black curve) with the same value of $\gamma$ as in Fig.~\ref{figure2}(c). Unlike the mobile EL emission resulting from a recombination of electron-hole pairs in ambipolar transport~\cite{avouris04}, the stationary nature of the EL emission and the gate-dependent EL emission intensity in our short channel SWNTFET devices strongly suggest that impact excitation is responsible for light emission. In contrast, phonon-assisted relaxation would lead to two very weak emission bands centered at around 950~nm and 700~nm and can therefore be excluded based on the observed EL spectrum~\cite{essig10}.

\begin{figure}[t]
\includegraphics[width=8.5cm]{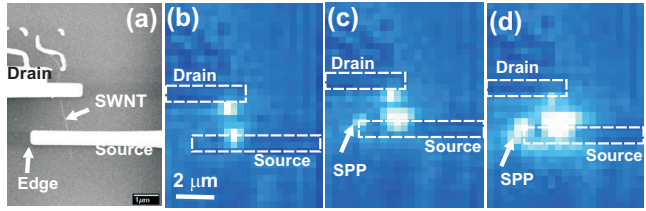}
\caption{\label{figure3} (Color online). (a) SEM image of a $\sim$1.8~$\mu$m-long SWNTFET. (b), (c) and (d) are overlaid electroluminescence images with $V_{GS}$=-9~V, -10~V, -11~V, respectively at $V_{DS}$=16~V. Scattering of light from the edge of the source electrode as shown in (c) and (d) demonstrates the excitation of SPP by the EL of the device.}
\end{figure}

We fabricated longer channel SWNTFET with specifically designed electrode geometries (Fig.~\ref{figure3} (a)), to demonstrate that the EL emission of the channel can be coupled to running SPPs in the electrical terminals. The device was driven at $\sim$1~$\mu$A drain current continuously for 30~s to acquire EL emission images  displayed in Figs.~\ref{figure3} (b) to (d) for different $V_{GS}$. EL from the device is observed at both source-nanotube and drain-nanotube contacts for $V_{DS}$=16~V and $V_{GS}$=-9~V (Fig.~\ref{figure3} (b)). The emission intensity at the source-nanotube contact increases with gate bias. Interestingly, we observe light emitted from the edge of the source electrode separated from the nanotube by about 2~$\mu$m at $V_{GS}$ of -10~V (Fig.~\ref{figure3}(c)). The intensity of light at the extremity is related to the intensity of the electroluminescent activity at the nanotube contact. Light emitted at the end facet of the electrode is a strong indication supporting the excitation of surface plasmon propagating in the source terminal. The SPP is developing in the electrode that also serves as a metal waveguide~\cite{weeber01PRB}. At the extremity, the SPP is scattered and is partially converted to detectable photons~\cite{vandorpe10,brongersma12NL,Berthelot:12}. 

\begin{figure}[t]
\includegraphics[width=8.5cm]{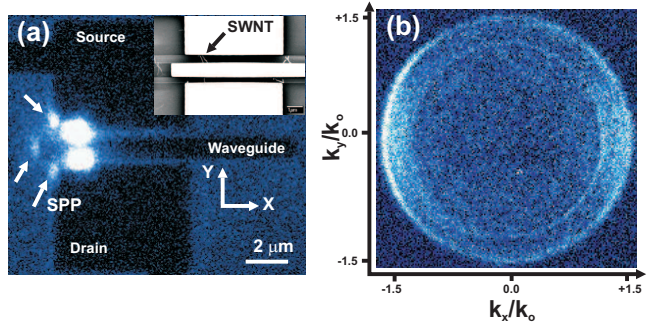}
\caption{\label{figure4} (Color online). (a) EL-coupled SPP launched in the source, drain and waveguide electrodes for $V_{GS}$=-5~V and $V_{DS}$=10~V and an integration time of 120~s. Inset: SEM image of a SWNTFET integrated with plasmonic compatible waveguide placed on top of the nanotube channel. (b) Fourier space image of the corresponding EL-coupled SPP shown in (a).}
\end{figure}

EL emission in SWNTs is known to be polarized along the axis of the nanotube~\cite{avourisscience03}. In the device shown in Fig.~\ref{figure3}(a), the nanotube is oriented at 75$^{\circ}$ with respect to the source electrode. Thus, there is small component of the electric field oriented along the main axis of the source electrode that can couple to a SPP. The intensity ratio of scattered SPPs to EL is 0.32 in both Fig.~\ref{figure3}(c) and Fig.~\ref{figure3}(d). To realize an electrical excitation of SPP in a bias-free electrode, we fabricated a SWNTFET by placing an additional receiving SPP waveguide on top of the nanotube channel. The excitation of SPP in this integrated device was achieved at $V_{DS}$=10~V and $V_{GS}$=-5~V. Figure~\ref{figure4}(a) shows the EL activity from the device. A strong response at the nanotube contacts is observed again concomitant to a weaker signal appearing at the left edges of the metal leads and of the bias-free waveguide. This scattered light emitted from the metal pads further confirm that SPPs can be excited with an electroluminescent plasmonic-compatible FET. 

To determine the nature of the excited SPP mode, we theoretically calculate for the effective index of the mode and its propagation length for our device geometry and emission wavelength. The dielectric constants for SiO$_{2}$, Ti, Au and air at 980~nm wavelength are $\epsilon_{SiO_2}$=2.25, $\epsilon_{Ti}$= 0.3+i21.5, $\epsilon_{Au}$= -39.9+i2.7, and $\epsilon_{air}$=1, respectively. The effective index and propagation length for a SPP mode at the Au/air interface are estimated at 1.01 and 40~$\mu$m respectively. For the Au thickness considered here, this SPP mode would leak part of its energy in the substrate and should be detected as leakages~\cite{Berthelot:12}. There is however no evidence of the presence of this leaky mode in the EL images presented in Fig.~\ref{figure3} and Fig.~\ref{figure4}. The propagation of a leaky SPP mode is limited by the width of the waveguide. The cutoff condition is approximately equal to the free space wavelength of the guided SPP signal~\cite{weeber01PRB,Berthelot:12,zia06}. In our configuration the width of the electrodes are below cutoff and the leaky mode is therefore vanishing. The second relevant SPP mode in the present geometry is a bound mode developing at the glass/metal interface. Its effective index and propagation length are 1.56 and 3~$\mu$m, respectively. We hypothesize that the SPP excited in the device configuration presented here is the bound mode present at the glass-metal interface. To confirm this assumption, we recorded the momentum-space (Fourier plane image)~\cite{Berthelot:12,massenot07} radiation pattern of the EL emission displayed in Fig.~\ref{figure4}(b). The angular detection is limited by the objective N.A. at 1.49. The critical angle at the glass/air interface is readily recognized (N.A=1.0). There is no indication of a leaky propagating SPP mode in this wave-vector distribution~\cite{Berthelot:12}. Instead, most of the light is emitted in a two-lobe pattern symmetric with respect to the reciprocal $k_x/k_o$ axis. Since the EL intensity at the source and drain contacts are the predominant signals in Fig.~\ref{figure4}(b), the recorded momentum distribution essentially reflects the dipolar-like emission pattern of the nanotube~\cite{hartmann12}. A maximum of intensity is observed at the outer rim for negative values of $k_x/k_o$. These observations are consistent with our original assumption.  SPP propagate along the waveguided and are scattered out at the left extremities of the electrodes. Since these edges are located at 1.5~$\mu$m to 2~$\mu$m from the CNT channel, a distance smaller than the propagation length of the bound SPP mode, a substantial fraction of SPPs is able to reach them. Scattering of the SPP by the sharp terminations is occurring at large wavevectors explaining the rise of intensity near the angular detection limit at -1.49. The mode is probably excited in both directions but, because the right extremities of the electrodes are located more than 6~$\mu$m away from the EL emission site, the plasmon looses most of its energy before reaching the edges and remains undetected.

In summary, we have demonstrated the use of a CNT-based opto-electronic device to produce an integrated electrical source of SPPs. Light emission in the devices is generated by a radiative relaxation of excitons formed by impact excitation. EL is partly coupled to bound SPPs that are conjointly transported with the electrical command in the SWNFET plasmonic-compatible terminals. In our devices, the coupling yield was limited by the relative orientation of the nanotube with respect to the electrode geometry; a limitation that can be addressed by careful designing. The electrodes used in this work only served as receptacles for the SPPs. Additional SPP routing and filtering functions could also be easily implemented by fine structuring the electrodes.  Finally, the transfer characteristics of SWNTFETs can be electrically modulated up to the MHz regime. Since the electroluminescent response and the SPP signal directly depend on the biasing condition, the device presented here could also serve to electrically modulate the surface plasmon intensity, a prerequisite to encode optical information in such a platform.

This work was funded partially by the NanoSci E+ program under grant E$^{2}$-PLAS (ANR-08-NSCI-007), the regional council of Burgundy with the program PARI, the Labex ACTION, the DFG through HA4405/5-1 and the Nanosystems Initiative Munich (NIM).

$^*$ Corresponding author: alexandre.bouhelier@u-bourgogne.fr


\begin{thebibliography}{35}
\expandafter\ifx\csname natexlab\endcsname\relax\def\natexlab#1{#1}\fi
\expandafter\ifx\csname bibnamefont\endcsname\relax
  \def\bibnamefont#1{#1}\fi
\expandafter\ifx\csname bibfnamefont\endcsname\relax
  \def\bibfnamefont#1{#1}\fi
\expandafter\ifx\csname citenamefont\endcsname\relax
  \def\citenamefont#1{#1}\fi
\expandafter\ifx\csname url\endcsname\relax
  \def\url#1{\texttt{#1}}\fi
\expandafter\ifx\csname urlprefix\endcsname\relax\def\urlprefix{URL }\fi
\providecommand{\bibinfo}[2]{#2}
\providecommand{\eprint}[2][]{\url{#2}}

\bibitem[{\citenamefont{Barnes et~al.}(2003)\citenamefont{Barnes, Dereux, and
  Ebbesen}}]{barnes03}
\bibinfo{author}{\bibfnamefont{W.~L.} \bibnamefont{Barnes}},
  \bibinfo{author}{\bibfnamefont{A.}~\bibnamefont{Dereux}}, \bibnamefont{and}
  \bibinfo{author}{\bibfnamefont{T.~W.} \bibnamefont{Ebbesen}},
  \bibinfo{journal}{Nature} \textbf{\bibinfo{volume}{424}},
  \bibinfo{pages}{824} (\bibinfo{year}{2003}).

\bibitem[{\citenamefont{Zia et~al.}(2006{\natexlab{a}})\citenamefont{Zia,
  Schuller, Chandran, and Brongersma}}]{brongersma06}
\bibinfo{author}{\bibfnamefont{R.}~\bibnamefont{Zia}},
  \bibinfo{author}{\bibfnamefont{J.~A.} \bibnamefont{Schuller}},
  \bibinfo{author}{\bibfnamefont{A.}~\bibnamefont{Chandran}}, \bibnamefont{and}
  \bibinfo{author}{\bibfnamefont{M.~L.} \bibnamefont{Brongersma}},
  \bibinfo{journal}{Materials Today} \textbf{\bibinfo{volume}{9}},
  \bibinfo{pages}{20} (\bibinfo{year}{2006}{\natexlab{a}}).

\bibitem[{\citenamefont{Ozbay}(2006)}]{ozbay06}
\bibinfo{author}{\bibfnamefont{E.}~\bibnamefont{Ozbay}},
  \bibinfo{journal}{Science} \textbf{\bibinfo{volume}{311}},
  \bibinfo{pages}{189} (\bibinfo{year}{2006}).

\bibitem[{\citenamefont{Ebbesen et~al.}(2008)\citenamefont{Ebbesen, Genet, and
  Bozhevolnyi}}]{ebbesen08}
\bibinfo{author}{\bibfnamefont{T.~W.} \bibnamefont{Ebbesen}},
  \bibinfo{author}{\bibfnamefont{C.}~\bibnamefont{Genet}}, \bibnamefont{and}
  \bibinfo{author}{\bibfnamefont{S.~I.} \bibnamefont{Bozhevolnyi}},
  \bibinfo{journal}{Physics Today} \textbf{\bibinfo{volume}{61}},
  \bibinfo{pages}{44} (\bibinfo{year}{2008}).

\bibitem[{\citenamefont{Dragoman and Dragoman}(2008)}]{DragomanPQE08}
\bibinfo{author}{\bibfnamefont{M.}~\bibnamefont{Dragoman}} \bibnamefont{and}
  \bibinfo{author}{\bibfnamefont{D.}~\bibnamefont{Dragoman}},
  \bibinfo{journal}{Prog. Quant. Electr.} \textbf{\bibinfo{volume}{32}},
  \bibinfo{pages}{1} (\bibinfo{year}{2008}).

\bibitem[{\citenamefont{Bouhelier and Wiederrecht}(2005)}]{bouhelier05OL}
\bibinfo{author}{\bibfnamefont{A.}~\bibnamefont{Bouhelier}} \bibnamefont{and}
  \bibinfo{author}{\bibfnamefont{G.~P.} \bibnamefont{Wiederrecht}},
  \bibinfo{journal}{Opt. Lett.} \textbf{\bibinfo{volume}{30}},
  \bibinfo{pages}{884} (\bibinfo{year}{2005}).

\bibitem[{\citenamefont{Hohenau et~al.}(2005)\citenamefont{Hohenau, Krenn,
  Stepanov, Drezet, Ditlbacher, Steinberger, Leitner, and
  Aussenegg}}]{Hohenau05}
\bibinfo{author}{\bibfnamefont{A.}~\bibnamefont{Hohenau}},
  \bibinfo{author}{\bibfnamefont{J.~R.} \bibnamefont{Krenn}},
  \bibinfo{author}{\bibfnamefont{A.~L.} \bibnamefont{Stepanov}},
  \bibinfo{author}{\bibfnamefont{A.}~\bibnamefont{Drezet}},
  \bibinfo{author}{\bibfnamefont{H.}~\bibnamefont{Ditlbacher}},
  \bibinfo{author}{\bibfnamefont{B.}~\bibnamefont{Steinberger}},
  \bibinfo{author}{\bibfnamefont{A.}~\bibnamefont{Leitner}}, \bibnamefont{and}
  \bibinfo{author}{\bibfnamefont{F.~R.} \bibnamefont{Aussenegg}},
  \bibinfo{journal}{Opt. Lett.} \textbf{\bibinfo{volume}{30}},
  \bibinfo{pages}{893} (\bibinfo{year}{2005}).

\bibitem[{\citenamefont{Walters et~al.}(2010)\citenamefont{Walters, van Loon,
  Brunets, Schmitz, and Polman}}]{polman10}
\bibinfo{author}{\bibfnamefont{R.~J.} \bibnamefont{Walters}},
  \bibinfo{author}{\bibfnamefont{R.~V.~A.} \bibnamefont{van Loon}},
  \bibinfo{author}{\bibfnamefont{I.}~\bibnamefont{Brunets}},
  \bibinfo{author}{\bibfnamefont{J.}~\bibnamefont{Schmitz}}, \bibnamefont{and}
  \bibinfo{author}{\bibfnamefont{A.}~\bibnamefont{Polman}},
  \bibinfo{journal}{Nat. Mat.} \textbf{\bibinfo{volume}{9}},
  \bibinfo{pages}{21} (\bibinfo{year}{2010}).

\bibitem[{\citenamefont{Neutens et~al.}(2010)\citenamefont{Neutens, Lagae,
  Borghs, and Van~Dorpe}}]{vandorpe10}
\bibinfo{author}{\bibfnamefont{P.}~\bibnamefont{Neutens}},
  \bibinfo{author}{\bibfnamefont{L.}~\bibnamefont{Lagae}},
  \bibinfo{author}{\bibfnamefont{G.}~\bibnamefont{Borghs}}, \bibnamefont{and}
  \bibinfo{author}{\bibfnamefont{P.}~\bibnamefont{Van~Dorpe}},
  \bibinfo{journal}{Nano Lett.} \textbf{\bibinfo{volume}{10}},
  \bibinfo{pages}{1429} (\bibinfo{year}{2010}).

\bibitem[{\citenamefont{Fan et~al.}(2012)\citenamefont{Fan, Colombo, Huang,
  Krogstrup, Nygård, Fontcuberta~i Morral, and Brongersma}}]{brongersma12NL}
\bibinfo{author}{\bibfnamefont{P.}~\bibnamefont{Fan}},
  \bibinfo{author}{\bibfnamefont{C.}~\bibnamefont{Colombo}},
  \bibinfo{author}{\bibfnamefont{K.~C.~Y.} \bibnamefont{Huang}},
  \bibinfo{author}{\bibfnamefont{P.}~\bibnamefont{Krogstrup}},
  \bibinfo{author}{\bibfnamefont{J.}~\bibnamefont{Nygård}},
  \bibinfo{author}{\bibfnamefont{A.}~\bibnamefont{Fontcuberta~i Morral}},
  \bibnamefont{and} \bibinfo{author}{\bibfnamefont{M.~L.}
  \bibnamefont{Brongersma}}, \bibinfo{journal}{Nano Letters}
  \textbf{\bibinfo{volume}{12}}, \bibinfo{pages}{4943} (\bibinfo{year}{2012}).

\bibitem[{\citenamefont{Bharadwaj et~al.}(2011)\citenamefont{Bharadwaj,
  Bouhelier, and Novotny}}]{Bharadwaj11}
\bibinfo{author}{\bibfnamefont{P.}~\bibnamefont{Bharadwaj}},
  \bibinfo{author}{\bibfnamefont{A.}~\bibnamefont{Bouhelier}},
  \bibnamefont{and} \bibinfo{author}{\bibfnamefont{L.}~\bibnamefont{Novotny}},
  \bibinfo{journal}{Phys. Rev. Lett.} \textbf{\bibinfo{volume}{106}},
  \bibinfo{pages}{226802} (\bibinfo{year}{2011}).

\bibitem[{\citenamefont{Koller et~al.}(2008)\citenamefont{Koller, Hohenau,
  Ditlbacher, Galler, Reil, Aussenegg, Leitner, List, and Krenn}}]{krenn08}
\bibinfo{author}{\bibfnamefont{D.}~\bibnamefont{Koller}},
  \bibinfo{author}{\bibfnamefont{A.}~\bibnamefont{Hohenau}},
  \bibinfo{author}{\bibfnamefont{H.}~\bibnamefont{Ditlbacher}},
  \bibinfo{author}{\bibfnamefont{N.}~\bibnamefont{Galler}},
  \bibinfo{author}{\bibfnamefont{F.}~\bibnamefont{Reil}},
  \bibinfo{author}{\bibfnamefont{F.}~\bibnamefont{Aussenegg}},
  \bibinfo{author}{\bibfnamefont{A.}~\bibnamefont{Leitner}},
  \bibinfo{author}{\bibfnamefont{E.}~\bibnamefont{List}}, \bibnamefont{and}
  \bibinfo{author}{\bibfnamefont{J.}~\bibnamefont{Krenn}},
  \bibinfo{journal}{Nature. Phot.} \textbf{\bibinfo{volume}{2}},
  \bibinfo{pages}{684} (\bibinfo{year}{2008}).

\bibitem[{\citenamefont{Avouris et~al.}(2008)\citenamefont{Avouris, Freitag,
  and Perebeinos}}]{avouris08}
\bibinfo{author}{\bibfnamefont{P.}~\bibnamefont{Avouris}},
  \bibinfo{author}{\bibfnamefont{M.}~\bibnamefont{Freitag}}, \bibnamefont{and}
  \bibinfo{author}{\bibfnamefont{V.}~\bibnamefont{Perebeinos}},
  \bibinfo{journal}{Nat. Photon.} \textbf{\bibinfo{volume}{2}},
  \bibinfo{pages}{341} (\bibinfo{year}{2008}).

\bibitem[{\citenamefont{Chen et~al.}(2005)\citenamefont{Chen, Perebeinos,
  Freitag, Tsang, Fu, Liu, and Avouris}}]{avouris05science}
\bibinfo{author}{\bibfnamefont{J.}~\bibnamefont{Chen}},
  \bibinfo{author}{\bibfnamefont{V.}~\bibnamefont{Perebeinos}},
  \bibinfo{author}{\bibfnamefont{M.}~\bibnamefont{Freitag}},
  \bibinfo{author}{\bibfnamefont{J.}~\bibnamefont{Tsang}},
  \bibinfo{author}{\bibfnamefont{Q.}~\bibnamefont{Fu}},
  \bibinfo{author}{\bibfnamefont{J.}~\bibnamefont{Liu}}, \bibnamefont{and}
  \bibinfo{author}{\bibfnamefont{P.}~\bibnamefont{Avouris}},
  \bibinfo{journal}{Science} \textbf{\bibinfo{volume}{310}},
  \bibinfo{pages}{1171} (\bibinfo{year}{2005}).

\bibitem[{\citenamefont{Freitag
  et~al.}(2004{\natexlab{a}})\citenamefont{Freitag, Chen, Tersoff, Tsang, Fu,
  Liu, and Avouris}}]{avouris04}
\bibinfo{author}{\bibfnamefont{M.}~\bibnamefont{Freitag}},
  \bibinfo{author}{\bibfnamefont{J.}~\bibnamefont{Chen}},
  \bibinfo{author}{\bibfnamefont{J.}~\bibnamefont{Tersoff}},
  \bibinfo{author}{\bibfnamefont{J.~C.} \bibnamefont{Tsang}},
  \bibinfo{author}{\bibfnamefont{Q.}~\bibnamefont{Fu}},
  \bibinfo{author}{\bibfnamefont{J.}~\bibnamefont{Liu}}, \bibnamefont{and}
  \bibinfo{author}{\bibfnamefont{P.}~\bibnamefont{Avouris}},
  \bibinfo{journal}{Phys. Rev. Lett.} \textbf{\bibinfo{volume}{93}},
  \bibinfo{pages}{076803} (\bibinfo{year}{2004}{\natexlab{a}}).

\bibitem[{\citenamefont{Essig et~al.}(2010)\citenamefont{Essig, Marquardt,
  Vijayaraghavan, Ganzhorn, Dehm, Hennrich, Ou, Green, Sciascia, Bonaccorso
  et~al.}}]{essig10}
\bibinfo{author}{\bibfnamefont{S.}~\bibnamefont{Essig}},
  \bibinfo{author}{\bibfnamefont{C.~W.} \bibnamefont{Marquardt}},
  \bibinfo{author}{\bibfnamefont{A.}~\bibnamefont{Vijayaraghavan}},
  \bibinfo{author}{\bibfnamefont{M.}~\bibnamefont{Ganzhorn}},
  \bibinfo{author}{\bibfnamefont{S.}~\bibnamefont{Dehm}},
  \bibinfo{author}{\bibfnamefont{F.}~\bibnamefont{Hennrich}},
  \bibinfo{author}{\bibfnamefont{F.}~\bibnamefont{Ou}},
  \bibinfo{author}{\bibfnamefont{A.~A.} \bibnamefont{Green}},
  \bibinfo{author}{\bibfnamefont{C.}~\bibnamefont{Sciascia}},
  \bibinfo{author}{\bibfnamefont{F.}~\bibnamefont{Bonaccorso}},
  \bibnamefont{et~al.}, \bibinfo{journal}{Nano Lett.}
  \textbf{\bibinfo{volume}{10}}, \bibinfo{pages}{1589} (\bibinfo{year}{2010}).

\bibitem[{\citenamefont{Freitag
  et~al.}(2004{\natexlab{b}})\citenamefont{Freitag, Perebeinos, Chen, Stein,
  Tsang, Misewich, Martel, and Avouris}}]{freitag04}
\bibinfo{author}{\bibfnamefont{M.}~\bibnamefont{Freitag}},
  \bibinfo{author}{\bibfnamefont{V.}~\bibnamefont{Perebeinos}},
  \bibinfo{author}{\bibfnamefont{J.}~\bibnamefont{Chen}},
  \bibinfo{author}{\bibfnamefont{A.}~\bibnamefont{Stein}},
  \bibinfo{author}{\bibfnamefont{J.~C.} \bibnamefont{Tsang}},
  \bibinfo{author}{\bibfnamefont{J.~A.} \bibnamefont{Misewich}},
  \bibinfo{author}{\bibfnamefont{R.}~\bibnamefont{Martel}}, \bibnamefont{and}
  \bibinfo{author}{\bibfnamefont{P.}~\bibnamefont{Avouris}},
  \bibinfo{journal}{Nano Letters} \textbf{\bibinfo{volume}{4}},
  \bibinfo{pages}{1063} (\bibinfo{year}{2004}{\natexlab{b}}).

\bibitem[{\citenamefont{Rao et~al.}(1997)\citenamefont{Rao, Richter, Bandow,
  Chase, Eklund, Williams, Fang, Subbaswamy, Menon, Thess et~al.}}]{Rao97}
\bibinfo{author}{\bibfnamefont{A.~M.} \bibnamefont{Rao}},
  \bibinfo{author}{\bibfnamefont{E.}~\bibnamefont{Richter}},
  \bibinfo{author}{\bibfnamefont{S.}~\bibnamefont{Bandow}},
  \bibinfo{author}{\bibfnamefont{B.}~\bibnamefont{Chase}},
  \bibinfo{author}{\bibfnamefont{P.~C.} \bibnamefont{Eklund}},
  \bibinfo{author}{\bibfnamefont{K.~A.} \bibnamefont{Williams}},
  \bibinfo{author}{\bibfnamefont{S.}~\bibnamefont{Fang}},
  \bibinfo{author}{\bibfnamefont{K.~R.} \bibnamefont{Subbaswamy}},
  \bibinfo{author}{\bibfnamefont{M.}~\bibnamefont{Menon}},
  \bibinfo{author}{\bibfnamefont{A.}~\bibnamefont{Thess}},
  \bibnamefont{et~al.}, \bibinfo{journal}{Science}
  \textbf{\bibinfo{volume}{275}}, \bibinfo{pages}{187} (\bibinfo{year}{1997}.

\bibitem[{\citenamefont{Hartmann et~al.}(2012)\citenamefont{Hartmann, Piredda,
  Berthelot, Colas~des Francs, Bouhelier, and Hartschuh}}]{hartmann12}
\bibinfo{author}{\bibfnamefont{N.}~\bibnamefont{Hartmann}},
  \bibinfo{author}{\bibfnamefont{G.}~\bibnamefont{Piredda}},
  \bibinfo{author}{\bibfnamefont{J.}~\bibnamefont{Berthelot}},
  \bibinfo{author}{\bibfnamefont{G.}~\bibnamefont{Colas~des Francs}},
  \bibinfo{author}{\bibfnamefont{A.}~\bibnamefont{Bouhelier}},
  \bibnamefont{and}
  \bibinfo{author}{\bibfnamefont{A.}~\bibnamefont{Hartschuh}},
  \bibinfo{journal}{Nano Letters} \textbf{\bibinfo{volume}{12}},
  \bibinfo{pages}{177} (\bibinfo{year}{2012}).

\bibitem[{\citenamefont{Rai et~al.}(2012)\citenamefont{Rai, Hartmann,
  Berthelot, des Francs, Hartschuh, and Bouhelier}}]{Rai:12}
\bibinfo{author}{\bibfnamefont{P.}~\bibnamefont{Rai}},
  \bibinfo{author}{\bibfnamefont{N.}~\bibnamefont{Hartmann}},
  \bibinfo{author}{\bibfnamefont{J.}~\bibnamefont{Berthelot}},
  \bibinfo{author}{\bibfnamefont{G.~C.} \bibnamefont{des Francs}},
  \bibinfo{author}{\bibfnamefont{A.}~\bibnamefont{Hartschuh}},
  \bibnamefont{and}
  \bibinfo{author}{\bibfnamefont{A.}~\bibnamefont{Bouhelier}},
  \bibinfo{journal}{Opt. Lett.} \textbf{\bibinfo{volume}{37}},
  \bibinfo{pages}{4711} (\bibinfo{year}{2012}).

\bibitem[{\citenamefont{Sumanasekera et~al.}(2000)\citenamefont{Sumanasekera,
  Adu, Fang, and Eklund}}]{Sumanasekera00}
\bibinfo{author}{\bibfnamefont{G.~U.} \bibnamefont{Sumanasekera}},
  \bibinfo{author}{\bibfnamefont{C.~K.~W.} \bibnamefont{Adu}},
  \bibinfo{author}{\bibfnamefont{S.}~\bibnamefont{Fang}}, \bibnamefont{and}
  \bibinfo{author}{\bibfnamefont{P.~C.} \bibnamefont{Eklund}},
  \bibinfo{journal}{Phys. Rev. Lett.} \textbf{\bibinfo{volume}{85}},
  \bibinfo{pages}{1096} (\bibinfo{year}{2000}).

\bibitem[{\citenamefont{Martel et~al.}(1998)\citenamefont{Martel, Schmidt,
  Shea, Hertel, and Avouris}}]{martel98}
\bibinfo{author}{\bibfnamefont{R.}~\bibnamefont{Martel}},
  \bibinfo{author}{\bibfnamefont{T.}~\bibnamefont{Schmidt}},
  \bibinfo{author}{\bibfnamefont{H.~R.} \bibnamefont{Shea}},
  \bibinfo{author}{\bibfnamefont{T.}~\bibnamefont{Hertel}}, \bibnamefont{and}
  \bibinfo{author}{\bibfnamefont{P.}~\bibnamefont{Avouris}},
  \bibinfo{journal}{Appl. Phys. Lett.} \textbf{\bibinfo{volume}{73}},
  \bibinfo{pages}{2447} (\bibinfo{year}{1998}).

\bibitem[{\citenamefont{Heinze et~al.}(2002)\citenamefont{Heinze, Tersoff,
  Martel, Derycke, Appenzeller, and Avouris}}]{avouris02}
\bibinfo{author}{\bibfnamefont{S.}~\bibnamefont{Heinze}},
  \bibinfo{author}{\bibfnamefont{J.}~\bibnamefont{Tersoff}},
  \bibinfo{author}{\bibfnamefont{R.}~\bibnamefont{Martel}},
  \bibinfo{author}{\bibfnamefont{V.}~\bibnamefont{Derycke}},
  \bibinfo{author}{\bibfnamefont{J.}~\bibnamefont{Appenzeller}},
  \bibnamefont{and} \bibinfo{author}{\bibfnamefont{P.}~\bibnamefont{Avouris}},
  \bibinfo{journal}{Phys. Rev. Lett.} \textbf{\bibinfo{volume}{89}},
  \bibinfo{pages}{106801} (\bibinfo{year}{2002}).

\bibitem[{\citenamefont{Freitag et~al.}(2009)\citenamefont{Freitag, Steiner,
  Naumov, Small, Bol, Perebeinos, and Avouris}}]{freitag09}
\bibinfo{author}{\bibfnamefont{M.}~\bibnamefont{Freitag}},
  \bibinfo{author}{\bibfnamefont{M.}~\bibnamefont{Steiner}},
  \bibinfo{author}{\bibfnamefont{A.}~\bibnamefont{Naumov}},
  \bibinfo{author}{\bibfnamefont{J.~P.} \bibnamefont{Small}},
  \bibinfo{author}{\bibfnamefont{A.~A.} \bibnamefont{Bol}},
  \bibinfo{author}{\bibfnamefont{V.}~\bibnamefont{Perebeinos}},
  \bibnamefont{and} \bibinfo{author}{\bibfnamefont{P.}~\bibnamefont{Avouris}},
  \bibinfo{journal}{ACS Nano} \textbf{\bibinfo{volume}{3}},
  \bibinfo{pages}{3744} (\bibinfo{year}{2009}).

\bibitem[{\citenamefont{Okuto and Crowell}(1972)}]{okuto72}
\bibinfo{author}{\bibfnamefont{Y.}~\bibnamefont{Okuto}} \bibnamefont{and}
  \bibinfo{author}{\bibfnamefont{C.~R.} \bibnamefont{Crowell}},
  \bibinfo{journal}{Phys. Rev. B} \textbf{\bibinfo{volume}{6}},
  \bibinfo{pages}{3076} (\bibinfo{year}{1972}).

\bibitem[{\citenamefont{Perebeinos et~al.}(2004)\citenamefont{Perebeinos,
  Tersoff, and Avouris}}]{perebeinos04}
\bibinfo{author}{\bibfnamefont{V.}~\bibnamefont{Perebeinos}},
  \bibinfo{author}{\bibfnamefont{J.}~\bibnamefont{Tersoff}}, \bibnamefont{and}
  \bibinfo{author}{\bibfnamefont{P.}~\bibnamefont{Avouris}},
  \bibinfo{journal}{Phys. Rev. Lett.} \textbf{\bibinfo{volume}{92}},
  \bibinfo{pages}{257402} (\bibinfo{year}{2004}).

\bibitem[{\citenamefont{Javey et~al.}(2004)\citenamefont{Javey, Guo, Paulsson,
  Wang, Mann, Lundstrom, and Dai}}]{javey04}
\bibinfo{author}{\bibfnamefont{A.}~\bibnamefont{Javey}},
  \bibinfo{author}{\bibfnamefont{J.}~\bibnamefont{Guo}},
  \bibinfo{author}{\bibfnamefont{M.}~\bibnamefont{Paulsson}},
  \bibinfo{author}{\bibfnamefont{Q.}~\bibnamefont{Wang}},
  \bibinfo{author}{\bibfnamefont{D.}~\bibnamefont{Mann}},
  \bibinfo{author}{\bibfnamefont{M.}~\bibnamefont{Lundstrom}},
  \bibnamefont{and} \bibinfo{author}{\bibfnamefont{H.}~\bibnamefont{Dai}},
  \bibinfo{journal}{Phys. Rev. Lett.} \textbf{\bibinfo{volume}{92}},
  \bibinfo{pages}{106804} (\bibinfo{year}{2004}).

\bibitem[{\citenamefont{Franklin et~al.}(2009)\citenamefont{Franklin, Sayer,
  Sands, Fisher, and Janes}}]{franklin09}
\bibinfo{author}{\bibfnamefont{A.~D.} \bibnamefont{Franklin}},
  \bibinfo{author}{\bibfnamefont{R.~A.} \bibnamefont{Sayer}},
  \bibinfo{author}{\bibfnamefont{T.~D.} \bibnamefont{Sands}},
  \bibinfo{author}{\bibfnamefont{T.~S.} \bibnamefont{Fisher}},
  \bibnamefont{and} \bibinfo{author}{\bibfnamefont{D.~B.} \bibnamefont{Janes}},
  \bibinfo{journal}{J. Vac. Sci. Technol. B} \textbf{\bibinfo{volume}{27}},
  \bibinfo{pages}{821} (\bibinfo{year}{2009}).

\bibitem[{\citenamefont{Fagan et~al.}(2007)\citenamefont{Fagan, Simpson, Landi,
  Richter, Mandelbaum, Bajpai, Ho, Raffaelle, Walker, Bauer et~al.}}]{fagan07}
\bibinfo{author}{\bibfnamefont{J.~A.} \bibnamefont{Fagan}},
  \bibinfo{author}{\bibfnamefont{J.~R.} \bibnamefont{Simpson}},
  \bibinfo{author}{\bibfnamefont{B.~J.} \bibnamefont{Landi}},
  \bibinfo{author}{\bibfnamefont{L.~J.} \bibnamefont{Richter}},
  \bibinfo{author}{\bibfnamefont{I.}~\bibnamefont{Mandelbaum}},
  \bibinfo{author}{\bibfnamefont{V.}~\bibnamefont{Bajpai}},
  \bibinfo{author}{\bibfnamefont{D.~L.} \bibnamefont{Ho}},
  \bibinfo{author}{\bibfnamefont{R.}~\bibnamefont{Raffaelle}},
  \bibinfo{author}{\bibfnamefont{A.~R.~H.} \bibnamefont{Walker}},
  \bibinfo{author}{\bibfnamefont{B.~J.} \bibnamefont{Bauer}},
  \bibnamefont{et~al.}, \bibinfo{journal}{Phys. Rev. Lett.}
  \textbf{\bibinfo{volume}{98}}, \bibinfo{pages}{147402}
  (\bibinfo{year}{2007}).

\bibitem[{\citenamefont{Anantram and L\'eonard}(2006)}]{Anantram06}
\bibinfo{author}{\bibfnamefont{M.~P.} \bibnamefont{Anantram}} \bibnamefont{and}
  \bibinfo{author}{\bibfnamefont{F.}~\bibnamefont{L\'eonard}},
  \bibinfo{journal}{Rep. Progr. Phys.} \textbf{\bibinfo{volume}{69}},
  \bibinfo{pages}{507} (\bibinfo{year}{2006}).

\bibitem[{\citenamefont{Weeber et~al.}(2001)\citenamefont{Weeber, Krenn,
  Dereux, Lamprecht, Lacroute, and Goudonnet}}]{weeber01PRB}
\bibinfo{author}{\bibfnamefont{J.-C.} \bibnamefont{Weeber}},
  \bibinfo{author}{\bibfnamefont{J.~R.} \bibnamefont{Krenn}},
  \bibinfo{author}{\bibfnamefont{A.}~\bibnamefont{Dereux}},
  \bibinfo{author}{\bibfnamefont{B.}~\bibnamefont{Lamprecht}},
  \bibinfo{author}{\bibfnamefont{Y.}~\bibnamefont{Lacroute}}, \bibnamefont{and}
  \bibinfo{author}{\bibfnamefont{J.~P.} \bibnamefont{Goudonnet}},
  \bibinfo{journal}{Phys. Rev. B} \textbf{\bibinfo{volume}{64}},
  \bibinfo{pages}{045411} (\bibinfo{year}{2001}).

\bibitem[{\citenamefont{Berthelot et~al.}(2012)\citenamefont{Berthelot,
  Tantussi, Rai, des Francs, Weeber, Dereux, Fuso, Allegrini, and
  Bouhelier}}]{Berthelot:12}
\bibinfo{author}{\bibfnamefont{J.}~\bibnamefont{Berthelot}},
  \bibinfo{author}{\bibfnamefont{F.}~\bibnamefont{Tantussi}},
  \bibinfo{author}{\bibfnamefont{P.}~\bibnamefont{Rai}},
  \bibinfo{author}{\bibfnamefont{G.~C.} \bibnamefont{des Francs}},
  \bibinfo{author}{\bibfnamefont{J.-C.} \bibnamefont{Weeber}},
  \bibinfo{author}{\bibfnamefont{A.}~\bibnamefont{Dereux}},
  \bibinfo{author}{\bibfnamefont{F.}~\bibnamefont{Fuso}},
  \bibinfo{author}{\bibfnamefont{M.}~\bibnamefont{Allegrini}},
  \bibnamefont{and}
  \bibinfo{author}{\bibfnamefont{A.}~\bibnamefont{Bouhelier}},
  \bibinfo{journal}{J. Opt. Soc. Am. B} \textbf{\bibinfo{volume}{29}},
  \bibinfo{pages}{226} (\bibinfo{year}{2012}).

\bibitem[{\citenamefont{Misewich et~al.}(2003)\citenamefont{Misewich, Martel,
  Avouris, Tsang, Heinze, and Tersoff}}]{avourisscience03}
\bibinfo{author}{\bibfnamefont{J.~A.} \bibnamefont{Misewich}},
  \bibinfo{author}{\bibfnamefont{R.}~\bibnamefont{Martel}},
  \bibinfo{author}{\bibfnamefont{P.}~\bibnamefont{Avouris}},
  \bibinfo{author}{\bibfnamefont{J.}~\bibnamefont{Tsang}},
  \bibinfo{author}{\bibfnamefont{S.}~\bibnamefont{Heinze}}, \bibnamefont{and}
  \bibinfo{author}{\bibfnamefont{J.}~\bibnamefont{Tersoff}},
  \bibinfo{journal}{Science} \textbf{\bibinfo{volume}{300}},
  \bibinfo{pages}{783} (\bibinfo{year}{2003}).

\bibitem[{\citenamefont{Zia et~al.}(2006{\natexlab{b}})\citenamefont{Zia,
  Schuller, and Brongersma}}]{zia06}
\bibinfo{author}{\bibfnamefont{R.}~\bibnamefont{Zia}},
  \bibinfo{author}{\bibfnamefont{J.~A.} \bibnamefont{Schuller}},
  \bibnamefont{and} \bibinfo{author}{\bibfnamefont{M.~L.}
  \bibnamefont{Brongersma}}, \bibinfo{journal}{Phys. Rev. B}
  \textbf{\bibinfo{volume}{74}}, \bibinfo{pages}{165415}
  (\bibinfo{year}{2006}{\natexlab{b}}).

\bibitem[{\citenamefont{Massenot et~al.}(2007)\citenamefont{Massenot,
  Grandidier, Bouhelier, {Colas des Francs}, Markey, Weeber, Dereux, Renger,
  Gonz\`alez, and Quidant}}]{massenot07}
\bibinfo{author}{\bibfnamefont{S.}~\bibnamefont{Massenot}},
  \bibinfo{author}{\bibfnamefont{J.}~\bibnamefont{Grandidier}},
  \bibinfo{author}{\bibfnamefont{A.}~\bibnamefont{Bouhelier}},
  \bibinfo{author}{\bibfnamefont{G.}~\bibnamefont{{Colas des Francs}}},
  \bibinfo{author}{\bibfnamefont{L.}~\bibnamefont{Markey}},
  \bibinfo{author}{\bibfnamefont{J.-C.} \bibnamefont{Weeber}},
  \bibinfo{author}{\bibfnamefont{A.}~\bibnamefont{Dereux}},
  \bibinfo{author}{\bibfnamefont{J.}~\bibnamefont{Renger}},
  \bibinfo{author}{\bibfnamefont{M.~U.} \bibnamefont{Gonz\`alez}},
  \bibnamefont{and} \bibinfo{author}{\bibfnamefont{R.}~\bibnamefont{Quidant}},
  \bibinfo{journal}{Appl. Phys. Lett.} \textbf{\bibinfo{volume}{91}},
  \bibinfo{pages}{243102} (\bibinfo{year}{2007}).

\end{thebibliography}
\end{document}